\documentclass[12pt,tightenlines,eqsecnum,floats,aps,amsmath,amssymb,nofootinbib,prd]{revtex4}

\def\an#1#2{\a{#1}{#2}{1}n}

\def\vc#1#2#3{{\,{#1}_{#2}\cdots\,{#1}_{#3}}}

\def\ind#1#2#3{^{\phantom{#1} #2}_{#1 \phantom{#2} #3}}

\def\a#1#2#3#4{{\,{#1}_{#3}{#2}_{#3}\cdots\,{#1}_{#4}{#2}_{#4}}}

\def\vk#1#2#3#4#5{\underline{\vc{#1}{#2}{#3}}\vc{#1}{#4}{#5}}

\def\OMEGA{\mbox{\large$\Omega$\normalsize}}

\begin{document}

\title{Covariance of the extended holonomy}

\author{Rodolfo Gambini$^{1}$, 
Jorge Pullin$^{2}$, Aureliano Skirzewski$^1$}
\affiliation {
1. Instituto de F\'{\i}sica, Facultad de Ciencias, 
Igu\'a 4225, esq. Mataojo, 11400 Montevideo, Uruguay. \\
2. Department of Physics and Astronomy, Louisiana State University,
Baton Rouge, LA 70803-4001}

\begin{abstract}
It has been pointed out that the holonomy of generic extended loops is
not gauge covariant. We show how to define a family of extended loops
for which previous criticism does not apply. We also give sufficient
conditions that extended loops must satisfy in order to yield 
covariant holonomies.  This makes a quantum representation for Yang--Mills
theories and gravity based on extended loops viable.
\end{abstract}
\maketitle

\section{Introduction}

The use of loop based variables to study gauge theories can be traced
all the way back to Faraday. In the context of Yang--Mills theory
holonomies have been widely used to analyze the quantization, both in
the continuum and in the lattice in gauge invariant terms. They have
also been used to base a complete quantum representation in terms of
loops, the loop representation, again both for Yang--Mills theories
and gravity (see \cite{GaPubook} and references therein).  Broadly
viewed, the holonomies can be seen as a way of providing test
functions against which to smear the fields of a theory. Because loops
are one dimensional objects, the resulting smearings tend to be
distributional in nature. This creates regularization problems when
operators act on the holonomies. In the context of gravity for
instance, where one wishes to have quantum states that are invariant
under diffeomorphisms, this has led to considering functions of
thickened or framed loops \cite{GaPugauss}. In the Yang--Mills
context, the definition of the inner product in the loop
representation would require summations over families of loops that
are not well defined. To deal with these issues the concept of
extended loops and the ensuing extended holonomies was introduced
\cite{extended,cmp}. The idea is to construct smearing functions that
share some of the properties of the smearings provided by loops, but
that are more general and have three-dimensional support. Expansions
formally similar to the ones used to define the non-Abelian holonomy
can be constructed in terms of those smeared functions to construct an extended holonomy.

It was shown \cite{Schilling}, however, that in spite of the formal analogy with the
case of loops, some convergence issues appeared in the expansion which
rendered the extended holonomy to be non-gauge covariant. A potential
solution was suggested \cite{DiBaGaGrPuJMP} to this problem by considering certain subsets
of extended holonomies. However, the solution was not entirely
satisfactory since the proposed subsets were ad-hoc in nature. In
spite of these difficulties the techniques attracted some attention in
the mathematical and particle physics literature \cite{mathematical}. Here we
would like to overcome those limitations by providing a generic
definition of the families of extended loops that yield properly
covariant holonomies. 

The structure of this paper is as follows. In
the next section we will review the concept of multitangents in
ordinary holonomies. In section 3 we discuss extended holonomies. In
section 4 we will propose the construction of extended loops that
yield covariant extended holonomies. Section 5 proposes an explicit 
construction of extended loops leading to covariant holonomies.  We end with a discussion.

\section{Ordinary holonomies and multitangents} 

The holonomy (whose trace is the {\em Wilson loop}) of a connection
$A_a$ is given by its path ordered exponential along a
loop\footnote{ It should be noted that in this context a ``loop'' is
  an equivalence class of curves that differ by retracings called
  ``trees''. All curves in the class yield the same holonomy and have
  the same multitangents.}. It can be rewritten as,
\begin{equation}
U_A(\gamma)=1+\sum_{n=1}^{\infty} \left(-i\right)^n \int dx^3_1\cdots dx^3_n
       A_{a_1}(x_1)\cdots A_{a_n}(x_n) T^{\vc a1n}(x_1,\ldots ,x_n,\gamma)
\label{holonomia}
\end{equation}
where $\gamma$ is a loop with a base point $o$ which we take as its origin and the loop
dependent multitangents $T$ are given by,
\begin{eqnarray}
T^{\vc a1n}(x_1,\ldots,x_n,\gamma) &=& \int_\gamma dy_n^{a_n}
   \int_0^{y_n}dy_{n-1}^{a_{n-1}}\cdots \int_0^{y_2}dy_{1}^{a_1}
   \delta (x_n-y_n)\cdots \delta(x_1-y_1) \nonumber
\\ & \label{X} \\
&=&\oint_\gamma dy_n^{a_n}\cdots \oint_\gamma dy_{1}^{a_1}
   \delta (x_n-y_n)\cdots \delta(x_1-y_1) \Theta_\gamma (0,y_1,\ldots,y_n)
\nonumber
\end{eqnarray}
with the multi Heaviside function $\Theta_\gamma(0,y_1,\ldots,y_n)$
ordering the points along the loop
starting at the origin and the Dirac deltas are three dimensional ones. These relations define the multitangents of
``rank" $n$. It will be convenient to introduce the notation
\begin{equation}
T^{\vc \mu 1n}(\gamma) = T^{\a ax1n}(\gamma) = T^{\vc a1n}(x_1,\ldots,x_n,\gamma) \;,
\end{equation}
with $\mu_i \equiv (a_ix_i)$, which suggests better the role played
by the x variables under diffeomorphisms \cite{extended,cmp}.

The multitangents satisfy a set of {\em algebraic identities}, which
follow directly from properties of the Heaviside function,
\begin{equation}
T^{\vk \mu 1k{k+1}n} \equiv
\sum_{P_k} T^{P_k(\vc \mu 1n)} = T^{\vc\mu 1k} \,T^{\vc\mu{k+1}n},
\label{VA}
\end{equation}
with the summation over all the permutations $P_k$ of the first $k$ of
the $\mu$ variables which
preserve the ordering of the ${ \mu_1, \ldots, \mu_k }$ and the ${
\mu_{k+1}, \ldots, \mu_n }$ among themselves.

They also satisfy a {\em differential constraint}, 
\begin{equation}
\frac {\partial \phantom{iii}} {\partial x_i^{a_i}} \,
T^{\a ax1i\,\ldots\,a_nx_n} =
 \bigl( \,\delta(x_i-x_{i-1}) - \delta(x_i-x_{i+1}) \,\bigr)
 T^{\a ax1{i-1}\,\a ax{i+1}n},
\label{dc}
\end{equation}
with $x_0$ and $x_{n+1}$ equal to the origin of the loop $o$.

The holonomy is gauge covariant, that is, if we consider a gauge
transformation $U$, 
\begin{equation}
  \label{eq:1}
  A_a \to A_a'=U A_a U^\dagger-\left(\partial_a U\right) U^\dagger,
\end{equation}
we have that,
\begin{equation}
  \label{eq:2}
  U_{A'}(\gamma) = U_o U_A(\gamma) U_o^\dagger
\end{equation}
Notice that the gauge transformation is a function of point and
$U_o$ is the gauge transformation evaluated at the origin of the loop.
The covariance of the holonomy follows from the path ordered nature along
the loop of the holonomy.  It is customary to take the loop origin at
infinity and the small gauge transformations (the ones connected to the
identity) as the identity at infinity. However, we will not make
assumptions about this in this paper.

A gauge field can be viewed as stemming 
from a representation of the group of loops in a Lie group
$G$. Every representation defines a connection up to gauge
transformations that leads to expansions for the holonomy that are
convergent (for a detailed discussion see \cite{GaTr83}). 

The multitangents transform as multivector densities under the subgroup of
coordinate transformations that leaves the base point $o$ fixed. That
is, if one has a transformation,
\begin{equation}
x^a \longrightarrow x'^a=D^a(x)
\end{equation}
then
\begin{equation}
T^{\an a{x'}}(DL) = \frac{\partial {x'}_1^{a_1}}{\partial x_1^{b_1}}
 \cdots \frac{\partial {x'}_n^{a_n}}{\partial x_n^{b_n}}
   \frac 1{J(x_1)}\cdots \frac 1{J(x_n)} T^{\an bx}(L) \,.
\label{ley}
\end{equation}
where $J$ is the Jacobian of the transformation. We will call objects
whose components transform in this way {\em multitensors}.

\section{Extended loops and holonomies}

Given a multitensor ${E} = (E^0, E^{\mu_1}, \cdots
,E^{\vc \mu 1n}, \cdots \, )$, where $E^0$ is a real constant that in
what follows we take equal to one, 
we can define an extended holonomy,
\begin{equation}
U_A(E) = 1+\sum_{n=1}^{\infty} \int \left(-i\right)^n dx_1\cdots dx_n
       A_{\mu_1}\cdots A_{\mu_n} E^{\vc\mu 1n} \,.
\label{holoE}
\end{equation}
with $E$ satisfying the differential and algebraic constraints. The
multitensors $E$ are a generalization of the multitangents. They have
a product,
\begin{equation}
  \label{eq:3}
  \left(E_1\times E_2\right)^{\mu_1\ldots\mu_n} = \sum_{i=0}^n
  E_1^{\mu_1\ldots \mu_i} E_2^{\mu_{i+1}\ldots \mu_n},
\end{equation}
that is related to the product of loops, which form a group
\cite{GaPubook}, 
\begin{equation}
  \label{eq:4}
  T\left(\gamma_1\circ\gamma_2\right)=T\left(\gamma_1\right)\times T\left(\gamma_2\right).
\end{equation}
The product is associative and satisfies the differential constraint. 

It is convenient to rewrite the generalized holonomy as,
\begin{equation}
U_A(E) = \sum_{n=0}^{\infty} 
       A_{\mu_1}\cdots A_{\mu_n} E^{\vc\mu 1n} ,
\end{equation}
where from now on $A_a(x)=-i A_a(x)$ and implicit in the sum are the integrals along space, with the term
with $n=0$ ($E$ with zero components, which we will call $E^0$) equal
to one. It is also useful to define the term containing $n$ powers of
the connection as,
\begin{equation}
  \label{eq:5}
U^{(n)}_A\left(E\right) =        A_{\mu_1}\cdots A_{\mu_n} E^{\vc\mu 1n}.
\end{equation}

Let us address the issue of gauge invariance. Consider an
infinitesimal gauge transformation,
\begin{equation}
  \label{eq:6}
  A^\Lambda=A + d \Lambda + \left[A,\Lambda\right]\,
\end{equation}
the terms in the sum transform as,
\begin{equation}
  \label{eq:7}
U_{A^\Lambda}^{(n)}=
U_{A}^{(n)}+\left[U_A^{(n-1)}(E),\Lambda_o\right]+f^{(n)})_{(A,\Lambda)}\left(E\right)-f^{(n-1)}_{(A,\Lambda)}\left(E\right),
\end{equation}
with $\Lambda_o$ the gauge transformation parameter evaluated at the origin of the
loop and,
\begin{equation}
  \label{eq:8}
  f^{(n)}_{(A,\Lambda)}\left(E\right) = \sum_k X^{\mu_1\ldots \mu_n}
  A_{\mu_1}\ldots A_{\mu_{k-1}}\left[A,\Lambda\right]_{\mu_k}
  A_{\mu_{k+1}}\ldots A_{\mu_n},
\end{equation}
and therefore,
\begin{equation}
  \label{eq:9}
  \sum_{n=0}^N U_{A^\Lambda}^{(n)}(E) = \sum_{n=0}^N
  U_A^{(n)}(E)+\left[\sum_{n=0}^N U_A^{(n)}(E),\Lambda_o\right]-\left[U_A^{(N)}(E),\Lambda_o\right]+f^{(N)}_{(A,\Lambda)}(E).
\end{equation}
 
If $U_A(E)$ converges, the next to last terms vanishes and the extended
holonomy is invariant if and only if $f_{(A,\Lambda)}^{(N)}(E)=0$ for
$N\to\infty$. It is not obvious that this holds for all $E$ and
$A$. In a nutshell, this was one of the points of the criticism in
\cite{Schilling}. 
The possibility of having extended loops for which $f(A,\Lambda)$ is
non-vanishing allows in principle \cite{Schilling} to find
counterexamples of extended loops that do not lead to gauge covariant
holonomies. In what follows we will show how to construct explicitly
extended loops that lead to gauge covariant holonomies and we will
establish sufficient conditions that ensure the covariance.

\section{Constructing extended invariants}

Let us see that one can define extended loops that lead to gauge
covariant extended holonomies. 
The starting point of the construction is the 
invariance for the case of loops, equation (\ref{eq:2}), and the
observation made in \cite{DiBaGaGrPuJMP}: one needs to restrict 
the type of extended loops considered, as we shall see in detail.  We will confine the
discussion to $SU(N)$ but it can be extended to other Lie groups. Let us define, as we did with the extended loop a
multitangent  $T(\gamma)$
with $T^0(\gamma)=1$ and introduce its product with a ``multiconnection''
(a product of connections),
\begin{equation}
  \label{eq:10}
  T(\gamma) \cdot A \equiv U_A(\gamma) = \sum_{n=0}^\infty T^{\mu_1\ldots
    \mu_n} A_{\mu_1}\ldots A_{\mu_n},
\end{equation}
and
\begin{equation}
  \label{eq:11}
  T(\gamma)\cdot A_g = U_{A_g}(\gamma)=U_g \left(T(\gamma)\cdot A\right) U^\dagger_g,
\end{equation}
where $A_g$ is the gauge transformed $A$ connection.
Let us define
\begin{equation}
  \label{eq:12}
  T(a,\gamma) \equiv (1-a){\cal I} +a T(\gamma),
\end{equation}
with ${\cal I}$ a multitensor that has as only non-vanishing component
${\cal I}^0=1$, that is, ${\cal I}^{\mu_1\ldots \mu_n}=0$ for $n\ne
0$. To put it another way,  $T^0(a,\gamma)=1$ and $T^{\mu_1\ldots\mu_n}(a,\gamma)=a T^{\mu_1\ldots\mu_n}(\gamma)$.
We therefore have that,
\begin{equation}
  \label{eq:13}
  T(a,\gamma)\cdot A = (1-a) I + a U_A(\gamma) \equiv U_A(a,\gamma),
\end{equation}
with $I$ the identity in the group, 
and for $a=1$ we have that $T(1,\gamma)\cdot A= U_A(1,\gamma) =
U_A(\gamma)$. Notice however, that $U(a,\gamma)$ is not an element of
the group if $a\ne 1$.

In order to define the family of extended loops that will lead to a
covariant holonomy, let us consider the expansion of the logarithm of $T$,
\begin{equation}
  \label{eq:14}
  F(a,\gamma) \equiv \ln\left(T(a,\gamma)\right)= \sum_{i=1}^\infty
  \frac{(-1)}{i} \left({\cal I} -T(a,\gamma)\right)^i,
\end{equation}
(the $i-th$ power is computed with the multitensor product introduced
before) and therefore,
\begin{equation}
  \label{eq:15}
F(a,\gamma)\cdot A= \sum_{i=1}^\infty \frac{(-1)}{i} \left(I-U_A(a,\gamma)\right)^i, 
\end{equation}
and we shall see that the series converges for $a$ sufficiently small.
Let $\| M
\|$ be the Frobenius norm of the matrix $M$ defined as,
\begin{equation}
  \label{eq:16}
  \| M\|\equiv \sqrt{{\rm Tr}\left( M^\dagger M\right)}=\sqrt{\sum_i \vert \lambda_i(M)\vert^2}
\end{equation}
with $\lambda_i$ the eigenvalues, for diagonalizable matrices. An
important property of the Frobenius norm is that it is submultiplicative
$\| A B\| \le \|A\| \|B\|$.

Let us consider the norm of $I-U_A(a,\gamma)$. If  we have that
\begin{equation}
  \label{eq:17}
  \| I-U_A(a,\gamma)\|< 1,
\end{equation}
given the submultiplicative property of the norm, this immediately
implies the expansion of the logarithm converges and therefore
$F(a,\gamma)$ is well defined. Evaluating,
\begin{eqnarray}
  \label{eq:18}
  \| I-U_A(a,\gamma)\| &=& \|a I-a U_A(\gamma)\|=\sqrt{{\rm
  Tr}\left(\biggl(a I-a U^\dagger_A(\gamma)\biggr)\biggl(a I-a U_A^{}(\gamma)\biggr)\right)}\nonumber\\
&=&\sqrt{{\rm Tr}\left(2 a^2 I -
    a^2\left(U^\dagger_A(\gamma)+U_A(\gamma)\right)\right)}\le 2
    a\sqrt{N}, 
\end{eqnarray}
which can be seen given the generic form of a unitary transformation
in $SU(N)$. Therefore $F(a,\gamma)\cdot A$ is well defined if $a <1/(2\sqrt{N})$.

Let us proceed to verify that the extended holonomy transforms
appropriately. 
Given that $\ln U_A(a,\gamma)= F(a,\gamma) \cdot A$ and taking into
account that,
\begin{eqnarray}
  \label{eq:19}
  U_{A_g}(a,\gamma)= U_{o} (1-a) U^\dagger_{o} + a U_{o} U_A(\gamma)
  U^\dagger_{o}=U_{o} U_A(a,\gamma) U^\dagger_{o},
\end{eqnarray}
where $U_{o}$ is the gauge transformation at the loop origin, and that,
\begin{equation}
  \label{eq:20}
  U_A(a,\gamma)=\exp\left(F(a,\gamma)\cdot A\right),
\end{equation}
it follows from (\ref{eq:15}) that
\begin{equation}
  \label{eq:21}
  F(a,\gamma)\cdot A_g = U_{o} F(a,\gamma) \cdot A U_{o}^\dagger.
\end{equation}

Equality (\ref{eq:20}) holds for $a< 1/(2\sqrt{N})$ and can be extended
analytically to the value $a=1$. To see this recall that the expansion
of the exponential of the logarithm of $x$ is well defined $\forall x$
even though the expansion of the logarithm is valid only for
$|x|<1$. The resulting series converges to $\ln(|x|)+i {\rm Arg}(x)$
with ${\rm Arg}(x)$ is the phase of $x$ in the interval $[-\pi,\pi]$. 

Evaluating at $a=1$ we have,
\begin{equation}
  \label{eq:22}
  U_A(1,\gamma) = U_A(\gamma),
\end{equation}
and 
\begin{equation}
  \label{eq:23}
  T(\gamma)=T(1,\gamma) = \exp\left(F(a,\gamma)\right)\vert_{a=1}.
\end{equation}
Note that $F(a,\gamma)\vert_{a=1}\cdot A$ belongs in the 
algebra (see appendix), its exponential is unitary and coincides with the 
principal logarithm of $U(a,\gamma)\vert_{a=1}$.

The analytic extension allows various generalizations of the concept
of loop that constitute extended loops for which the holonomy
transforms covariantly. 

For instance, starting from (\ref{eq:20}) and using (\ref{eq:21}) we can define a
gauge covariant real power of a holonomy,
\begin{equation}
  \label{eq:24}
  U_{A_g}(a,\gamma)^\lambda = \exp\left(\lambda F(a,\gamma)\cdot
    A_{g}\right) = \exp\left(U_{o} \lambda F(a,\gamma)\cdot A
    U^\dagger_{o}\right)=U_{o} U_A^\lambda(a,\gamma) U^\dagger_{o},
\end{equation}
and $U_A(a=1,\gamma)^\lambda$ is the principal branch of the real $\lambda$-th power of a
holonomy associated with,
\begin{equation}
  \label{eq:25}
  T^\lambda(\gamma)=\exp\left(\lambda F(a,\gamma)\right)\vert_{a=1},
\end{equation}
which is an example of an extended loop that leads to a covariant
holonomy. There are many examples of covariant extensions of loops. The
technique presented obviously includes ordinary loops. Real powers of
loops are clearly invertible and form a group, the associated
holonomies are unitary and gauge covariant.

The covariant extensions stem from observing that the analytic
extension,
\begin{equation}
  \label{eq:26}
  F_A(\gamma)\equiv F(a,\gamma)\cdot A\vert_{a=1},
\end{equation}
belongs to the $SU(N)$ algebra (see appendix) and is gauge covariant for all loops
$\gamma$. That implies that,
\begin{equation}
  \label{eq:27}
  F_A(\gamma_1)+F_A(\gamma_2), \lambda F_A(\gamma), \left[F_A(\gamma_1),F_A(\gamma_2)\right],
\end{equation}
lead, through exponentiation, to elements of the group that are gauge
invariant and define an extended loop algebra with their corresponding
extended holonomies. 

The idea that this algebra allows to define
smoothed loops can be confirmed considering a bi-parametric family
$\gamma(\alpha,\beta)$ of loops. The quantity,
\begin{equation}
  \label{eq:28}
\left.  \exp\left(\int d\alpha d\beta \lambda(\alpha,\beta)\left[F(a,\gamma(\alpha,\beta))\cdot
      A\right] \right)\right\vert_{a=1}=U_A(\gamma(\alpha,\beta),\lambda(\alpha,\beta))
\end{equation}
where $\lambda(\alpha,\beta)$ a suitable functional coefficient for each member
of the family is an example of smoothed loop. Summarizing, starting
from ordinary loops one can construct a large family of extended loops
that lead to covariant holonomies. The key observation is that
although
$F(a,\gamma)\vert_{a=1}$, given by (\ref{eq:14}) in general is not well defined,
since the convergence radius of the expansion of the exponential is
infinite, $\left[\exp\left(F(a,\gamma)\right)\right]\vert_{a=1}$ is. 

Summarizing, given $T(\gamma)= \exp\left(F(a,\gamma)\right)\vert_{a=1}$, we can
define a family of extended loops $E(\lambda F_\gamma)$,
\begin{equation}
  \label{eq:29}
  T^\lambda(\gamma) = \exp\left(\lambda F(a,\gamma)\right)\vert_{a=1}
  \equiv E(\lambda F_\gamma),
\end{equation}
and construct, 
\begin{equation}
  \label{eq:30}
  E\left(\lambda F_{\gamma_1}+\mu F_{\gamma_2}\right) = \exp\left(
    \lambda F(a,\gamma_1)+\mu F(a,\gamma_2)\right)\vert_{a=1},
\end{equation}
or $E(\int d\alpha d\beta \lambda(\alpha,\beta)
F_{\gamma(\alpha,\beta)})$, or 
$E\left(\left[F_{\gamma_1},F_{\gamma_2}\right]\right)$ and $E$ of 
multiple commutators defined analogously. All of them take the form,
\begin{equation}
  \label{eq:31}
  E=\left( E^0 = 1, E^{\mu_1}, E^{\mu_1,\,\mu_2}, \ldots, E^{\mu_1\ldots \mu_n}\right),
\end{equation}
but the $E^{\mu_1\ldots \mu_n}$ satisfy additional conditions to the
differential and algebraic constraints: they are exponentials of
$F$'s, as constructed above. 
The components of $E$ lead to series that converge to unitary and
gauge covariant transformations  and are examples of extensions that
satisfy equations (3.3) and (3.4) of \cite{Schilling}.

Notice that the $F$'s satisfy the differential constraint and a
simpler version of the algebraic constraint given by
$F(a,\gamma)^{\underline{\mu_1\dots\mu_k}\mu_{k+1}\dots
  \mu_n}\Big|_{a=1}=0$. This constraint is a key ingredient in the construction of
elements of the Lie group algebra \cite{cmp}.

To prove that the constraint is satisfied, let us consider the continuous binomial expansion
\begin{equation}
    T^\lambda(a,\alpha)=(\mathcal{I}(1-a)+a T(\alpha))^\lambda=\sum_{m=0}^\infty \binom{\lambda}{m} (1-a)^m a^{\lambda-m} T^{\lambda-m}(\alpha),
\end{equation}
with $\alpha$ a loop. 
Given that the multitangents satisfy the algebraic constraint\begin{equation}
    T^{\lambda-m}(\alpha)^{\underline{\mu_1\dots\mu_k}\mu_{k+1}\dots\mu_n} = T^{\lambda-m}(\alpha)^{\mu_1\dots\mu_k} T^{\lambda-m}(\alpha)^{\mu_{k+1}\dots\mu_n},
\end{equation}
differentiating the product in $T^\lambda(a,\alpha)^{\underline{\mu_1\dots\mu_k}\mu_{k+1}\dots\mu_n}$,
\begin{align}
    \frac{d\ }{d\lambda}T^\lambda(a,\alpha)^{\underline{\mu_1\dots\mu_k}\mu_{k+1}\dots\mu_n}&=\sum_{m=0}^\infty (1-a)^m\frac{d\ }{d\lambda}\Big(\binom{\lambda}{m}  a^{\lambda-m}\Big) T^{\lambda-m}(\alpha)^{\mu_1\dots\mu_k} T^{\lambda-m}(\alpha)^{\mu_{k+1}\dots\mu_n}\nonumber\\
    &+\sum_{m=0}^\infty \binom{\lambda}{m} (1-a)^m a^{\lambda-m} \frac{d\ }{d\lambda}T^{\lambda-m}(\alpha)^{\mu_1\dots\mu_k} T^{\lambda-m}(\alpha)^{\mu_{k+1}\dots\mu_n}\nonumber\\
    &+\sum_{m=0}^\infty \binom{\lambda}{m} (1-a)^m a^{\lambda-m} T^{\lambda-m}(\alpha)^{\mu_1\dots\mu_k} \frac{d\ }{d\lambda}T^{\lambda-m}(\alpha)^{\mu_{k+1}\dots\mu_n},
\end{align}
 where $T^{-m}(\alpha)\equiv T^{m}(\bar\alpha)$ \cite{cmp}. Evaluating the derivative at $\lambda=0$, we obtain the algebraic constraint of the generators of $T^\lambda(a,\alpha)$,
\begin{align}
    F(a,\alpha)^{\underline{\mu_1\dots\mu_k}\mu_{k+1}\dots\mu_n}=&\frac{d\ }{d\lambda}T^\lambda(a,\alpha)^{\underline{\mu_1\dots\mu_k}\mu_{k+1}\dots\mu_n}\Big|_{\lambda=0}\\
    =&\sum_{m=1}^\infty (\frac{1}{a}-1)^m \frac{(-1)^{m-1}}{m}(1+\gamma(m-1)) T^{m}(\bar\alpha)^{\mu_1\dots\mu_k} T^{m}(\bar\alpha)^{\mu_{k+1}\dots\mu_n}\nonumber\\ +&F(\alpha)^{\mu_1\dots\mu_k} \mathcal{I}^{\mu_{k+1}\dots\mu_n}\nonumber+\mathcal{I}^{\mu_1\dots\mu_k} F(\alpha)^{\mu_{k+1}\dots\mu_n},
\end{align} where  $\gamma$ is the Euler--Mascheroni constant.
Finally,  for $1<k<n$ and $a\rightarrow 1$,
$F(a,\gamma)^{\underline{\mu_1\dots\mu_k}\mu_{k+1}\dots
  \mu_n}\Big|_{a=1}=0$ since $\mathcal{I}^{\mu_1\dots\mu_n}=0$ for
$n\ge 1$.
This ensures that the product $F\cdot A$ is in the algebra (see appendix).
One can also demonstrate that the exponential of any quantity
satisfying the homogeneous algebraic constraint produces an object that satisfies the algebraic
constraint.

\section{An explicit characterization of extended loops leading to
  covariant holonomies}

Up to now we have followed a constructive process to identify extended
loops, either considering real powers of a loop or more general
constructions, always starting from ordinary loops. However, it is
convenient to have a notion of extended loops that lead to covariant
holonomies,  independent of their 
construction procedure . 

The non covariance of the extended holonomy discussed in
\cite{Schilling} was based on a two fold argument. One of them can be
solved by regularization: gauge transformations of extended holonomies
based on real powers of loop holonomies may appear to be non invariant
due to the appearance of different branches of the exponentiation of a
function by a real parameter.
The complex power function is a multi-valued function
The principal branch of the function is obtained by replacing $\ln(z)$ with the principal branch of the logarithm. 
If one adds to this observation that ordinary loops
lead to gauge invariant holonomies then this leads to extensions that
also yield gauge invariant holonomies as we showed in previous sections. 

The second problem regards the difficulty to prove that the gauge
transformation of the local connection transforms correctly in the
limit where the holonomy includes an infinite number of nodes (gauge
field insertions) in the extended loop. That is, the last term in
(\ref{eq:15}) vanishes. It is known that holonomies constructed in
terms of loops transform correctly and one can prove it as in
\cite{Giles}, by partitioning the loop $\gamma$ into a collection of
infinitesimal straight segments $S_{z_i}^{z_{i+1}}$ that form a
polygonal that, as you increase the number of segments approaches the
curve . For a large number $N+1$ of segments
\begin{eqnarray}
U_A(\gamma)=    {\mathcal{P}}\exp\Big\{\int_\gamma A_a(x)dx^a\Big\}&\simeq&
  \Big(1+(x_{N+1}^{a_{N}}-x_{N}^{a_N})A_{a_N}(x_N)+O((x_N-x_{N-1})^2)\Big)\dots\nonumber\\&&\times
  \Big(1+(x_1^{a_0}-x_0^{a_0})A_{a_0}(x_0)+O((x_1-x_0)^2)\Big), \label{8.1}
\end{eqnarray}
and we note that in the limit $N\to \infty$ the right hand side
reproduces the left hand side. 
We can identify the loop holonomy as an ordered product of
infinitesimal open path parallel transports \begin{equation}
    U_A(S_{x_i}^{x_{i+1}})=1+(x_{i+1}^{a_i}-x_{i}^{a_i})A_{a_i}(x_i)+O((x_{i+1}-x_{i})^2)
\end{equation} where it has been assumed that the distance between one point and the next is infinitesimal.
It can easily be proven that arbitrary gauge transformations act
as $$U_{A^g}[S_{x_i}^{x_{i+1}}]=U_g(x_i)U_{A}[S_{x_i}^{x_{i+1}}]U^\dagger_g(x_{i+1})+O\left((x_{i+1}-x_i)^2\right),$$
with $g$ the element of the group associated with the gauge transformation,
and we immediately get in the limit where the infinitesimal intervals
go to zero $$U_{A^g}[\gamma]=U_g(x_0))U_{A}[\gamma]U^\dagger_g(x_{0})),$$
where we have used $x_{N+1}=x_0$ and $U_g(x_0)=U_{o}$.

To understand how these results extend to the case of extended loops,
let us rewrite the above expression in terms of multitangents. To this
aim, it is convenient to partition space into cubes and consider the
set of cubes that are intersected by the path.  We consider a cubic
lattice characterized by a lattice size $l$, and substitute the curve
inside each cell by a straight line, entering through $y_{i}$ and
exiting through $y_{i+1}$, and composing the
straight lines into an $N+1$ sided polygonal. Substituting the tangent
vector
$dy^{a}\rightarrow y_{i+1}^{a}-y_{i}^{a}=\epsilon_i \hat u_i^a$, with
$\epsilon_i$ of order $l$ and $\hat{u}_i^a$ unit vectors, 
we get,
\begin{eqnarray}
&&T^{a_1\ldots a_n}\left(x_1,\ldots,x_n,\gamma\right)=
    \int_{\gamma} dy^{a_n}_n\int_0^{y_n} dy^{a_{n-1}}_{n-1}\cdots\int_0^{y_{2}} dy^{a_1}_1\delta(x_n-y_n)\cdots\delta(x_1-y_1)\nonumber\\
    &&\rightarrow T_N^{a_1x_1\ldots
       a_nx_n}(\gamma)
=\sum_{\begin{array}{c}
i_1<i_2\ldots< i_n \cr 0\le i_j\le N\end{array}} \epsilon_{i_1}\cdots \epsilon_{i_n}
       \hat{u}_{i_1}^{a_1}\cdots \hat{u}_{i_n}^{a_n}\delta(y_{i_1}
       -x_1) \cdots \delta(y_{i_n}-x_n),
\end{eqnarray}
where $y_{i_j}^a=\sum_{k=0}^{i_j-1} \epsilon_k {\hat{u}}^{a}+y_0^a$ . Notice that this expression is valid for $n\le N+1$ since the
inequalities in the sum cannot be satisfied if $n> N+1$, in that case
$T_N^{a_1x_1\ldots a_n x_n}=0$. 

Taking into account (\ref{8.1})  we have that,
\begin{equation}
  \label{eq:38}
  U_A(\gamma) = \lim_{N\to \infty} \left(1+  T^{\mu_1}_N A_{\mu_1} +
    \cdots +  T_N^{\mu_1 \ldots \mu_p} A_{\mu_1}\cdots
    A_{\mu_p}+\cdots + 
    T_N^{\mu_1\ldots \mu_{N+1}} A_{\mu_1} \cdots A_{\mu_{N+1}}\right)
\end{equation}
and the term,
\begin{equation}
  \label{eq:39}
    T_N^{\mu_1\ldots \mu_N} A_{\mu_1} \cdots A_{\mu_N}=O\left(\scriptstyle{\left(\frac{1}{N}\right)^N}\right),
\end{equation}
and this is true for any finite connection $A$. In particular for both
$A$ and its gauge transformed $A_g$. Let us note that the polygonal
multitangents $T_N$ satisfy the algebraic constraint and the
differential constraint up to higher order terms, 
\begin{eqnarray}
  \partial_{x^a_i} T_N^{a_1 x_1,\ldots, a_i x_i,\ldots, a_n x_n} &=&
  \Theta( N-n) \left[\delta(x_i-x_{i+1} ) -\delta(x_i-x_{i-1})\right]
  T_N^{a_1 x_1,\ldots, a_{i-1} x_{i-1},a_{i+1}x_{i+1} ,\ldots, a_n
    x_n}\nonumber\\ && + O\left(\left(x_{i+1}-x_{i-1}\right)^2\right).\label{diff41}
\end{eqnarray}

Since expression (\ref{eq:38}) is an explicit form of (\ref{8.1})
which is manifestly gauge invariant in the limit $N\to\infty$,
(\ref{eq:38}) is too. This can be verified directly observing that the
remainder term of $f^{(N)}_{(A,\Lambda)}(E)$ of (\ref{eq:9}) vanishes
due to (\ref{eq:39}) in what concerns the algebraic constraint 
\begin{equation}
T^{\vk \mu 1k{k+1}n}= T^{\vc\mu 1k} \,T^{\vc\mu{k+1}n},
\end{equation}
Notice, however, that the polygonal multitangents do not form a group
because $T_N \times T_M =T_{N+M+1}$.  However the inverse of $T_N$ is a
$T_N$ and the polygonal loops for arbitrary $N$ do form a group.

So the multitangent,
\begin{equation}
  \label{eq:41}
  T= \lim_{N\to \infty} T_N,
\end{equation}
where $T$ are the vectors associated with the multitangent in the
notation of (\ref{eq:31}) and,
\begin{equation}
  \label{eq:42}
  \lim_{N\to\infty} T_N^{\mu_1\ldots \mu_N} =0. 
\end{equation}

These properties of the multitangents can be directly generalized to
extended loops giving a sufficient criterion for the latter to lead to
a covariant holonomy: the extended loop must be given by a limit
$E=\lim_{N\to \infty} E_N$  with the $E_N$'s satisfying a differential constraint
(\ref{diff41}) and a 
condition like (\ref{eq:39}) that implies 
limit like (\ref{eq:42}).

It is immediate to show that the explicit constructions presented in
previous sections satisfy these conditions for a suitable definition
of the limits involved. Let us see this explicitly for the case of a
real power $\lambda$ of a loop. Let us define a polynomial approximation to the algebra element $F(a,\gamma)$,
\begin{equation}
  \label{eq:43}
  F(m,a,\gamma)=\sum_{i=1}^m \frac{-1}{i} \left({\cal I} -
    T_{m^4}(a,\gamma) \right)^i,
\end{equation}
with the extended loop,
\begin{equation}
  \label{eq:44}
  E_{m^4}(a,\gamma^\lambda) = \sum_{j=0}^m \frac{\left(\lambda F(m,a,\gamma)\right)^j}{j!}. 
\end{equation}
It is clear that this expression satisfies the previous
conditions. Noticing that the highest order term comes from $F(m,a,\gamma)^m$
and it takes the form,
\begin{equation}
  \label{eq:45}
  T_m(a,\gamma)^{\mu_1 \ldots \mu_m} 
\times   T_m(a,\gamma)^{\mu_{m+1} \ldots \mu_{2m} }
\times   T_m(a,\gamma)^{\mu_{2m+1} \ldots \mu_{3m} }
\times   T_m(a,\gamma)^{\mu_{3m+1} \ldots \mu_{4m} }
\end{equation}
which obviously goes to zero as $T_m^{\mu_1\ldots \mu_m}$ goes to zero.

\section{Conclusions}

We have shown how to generate large families of extended loops that
yield covariant holonomies. They include ordinary loops and real
powers of them, among others. The real powers of loops constitute a
group with associated holonomies that are unitary and gauge
invariant. The center of the idea is to construct extended loops using
the expansion of the logarithm of multitensors. Through an analytic
extension they can be shown to yield covariant holonomies.
We have also given sufficient conditions for extended loops that lead
to covariant holonomies and showed that the extended loops obtained
from the previous construction satisfy them.

 This opens
the possibility of using the ensuing extended loops to create extended
loop representations of interest for the non perturbative quantization of Yang--Mills
theories and potentially gravity. In the case of Yang--Mills theories
the use of extended loops could have advantages over the use of ordinary
loops when one wishes to define the inner product and the closure
relations, and for the renormalization of non-perturbative Schr\"odinger-like equations.

\section{Acknowledgments}
This work was supported in part by Grants NSF-PHY-1603630,
NSF-PHY-1903799, funds of the Hearne Institute for Theoretical
Physics, CCT-LSU, Pedeciba and Fondo Clemente Estable
FCE\_1\_2014\_1\_103803.

\section*{Appendix}
Let us show explicitly that $F(a,\gamma)\cdot A$ with $F$ satisfying the homogeneous
algebraic identity $F^{\vk \mu 1k{k+1}n} =0 $   is in the algebra.
To do that we use a technique developed in \cite{extended,cmp}. 
We define the matrix of delta functions, 
\begin{equation}
\delta\ind{}{\vc \mu 1n}{\vc \nu 1m} = \delta_{n,m} \delta^{\mu_1}_{\nu_1}
\cdots \delta^{\mu_n}_{\nu_n}
\end{equation}
and the vector (using the same notation as in (\ref{eq:31})),
\begin{equation}
{\bf\delta}_{\vc \nu 1i} \,=\, (\;0,\;
\delta\ind{}{\mu_1}{\vc \nu 1i}, \; \cdots ,
\delta\ind{}{\vc \mu 1n}{\vc \nu 1i}, \; \cdots
\,) \;\;.
\end{equation}
We define then a projector from generic multitangents to those
satisfying the homogeneous algebraic constraint, given by
\begin{equation}
\OMEGA\ind{}{\vc \mu 1n}{\vc \nu 1m} \equiv \frac{\delta_{n,m}}{m}
[\,[ \, \cdots \, [ \delta_{\nu_1} ,\delta_{\nu_2} ] , \, \cdots \, ],
\delta_{\nu_n}]^{\vc \mu 1n} \theta(m-1) \, + \, \delta_{m,1}
\delta^{\mu_1}_{\nu_1}
\end{equation}
where 
\begin{equation}
[\delta_{\nu_1},\delta_{\nu_2}]^{\mu_1\mu_2}=(\delta_{\nu_1}\delta_{\nu_2}-\delta_{\nu_2}\delta_{\nu_1})^{\mu_1\mu_2}=
\delta_{\nu_1}^{\mu_1}\delta_{\nu_2}^{\mu_2} -
\delta_{\nu_2}^{\mu_1}\delta_{\nu_1}^{\mu_2},
\end{equation}
and the $\theta$ function means that the terms is non-vanishing for
$m>1$.  From the above definition,
\begin{eqnarray}
  \label{eq:35}
   \Omega^{\vk \mu 1k{k+1}{n+1}}{}_{\nu_1\ldots \nu_{n+1}}&=& \frac{n}{n+1}\left(
\Omega^{\vk \mu 1k{k+1}n}{}_{\nu_1\ldots
  \nu_{n}}\delta^{\mu_{n+1}}_{\nu_{n+1}}\right.\nonumber\\
&+&
\Omega^{\vk \mu 1{k-1}{k+1}{n+1}}{}_{\nu_1\ldots
                                                              \nu_{n}}\delta^{\mu_{k}}_{\nu_{n+1}}\nonumber\\
&-&\!\!
\delta^{\mu_1}_{\nu_{n+1}}\Omega^{\vk \mu 2k{k+1}{n+1}}{}_{\nu_1\ldots
  \nu_{n}}\nonumber\\
&-&\left.
\delta^{\mu_{k+1}}_{\nu_{n+1}}\Omega^{\vk \mu 1{k}{k+2}{n+1}}{}_{\nu_1\ldots
                                                              \nu_{n}}\right),
\end{eqnarray}
we immediately have that if,
\begin{equation}
 \Omega^{\vk \mu 1l{l+1}{n}}{}_{\nu_1\ldots \nu_{n}}=0,\quad \forall
 l/1\le l<n,
\end{equation}
then,
\begin{equation}
 \Omega^{\vk \mu 1k{k+1}{n+1}}{}_{\nu_1\ldots \nu_{n+1}}=0,\quad \forall
k/1\le l<n+1.
\end{equation}

Given that $\Omega$ is a projector, one has that 
\begin{equation}
  \label{eq:36}
F \cdot A = \left(\Omega\cdot F\right) \cdot A = F\cdot \left(\Omega\cdot  A\right),
\end{equation}
and $\Omega \cdot A$ is in the algebra and $F \cdot A$ is too. This
can be seen in the following way,
\begin{eqnarray}
  \label{eq:37}
  \left(\Omega_{\nu_1\ldots \nu_n} \cdot A \right)_{\nu_1\ldots
  \nu_n}&=& \Omega^{\mu_1\ldots \mu_n}{}_{\nu_1\ldots \nu_n} A_{\mu_1}
  \cdots A_{\mu_n} \\&=&\frac{1}{n} \left(\Omega^{\mu_1\ldots
  \mu_{n-1}}{}_{\nu_1\ldots \nu_{n-1}} A_{\mu_1} \cdots
  A_{\mu_{n-1}}\cdots A_{\mu_{n-1}} A_{\nu_n}  \right.\nonumber\\
&&\left.-A_{\nu_n}
  \Omega^{\mu_2\ldots \mu_n}{}_{\nu_1\ldots \nu_{n-1}} A_{\mu_2}\cdots
  A_{\mu_n}\right)\\
&=& \frac{1}{n} \left[ \Omega_{\nu_1\ldots \nu_{n-1}}\cdot 
    A,A_{\nu_n}\right]\\
&&\vdots \nonumber\\
&=& \frac{1}{n!} \left[\ldots \left[A_{\nu_1},A_{\nu_2}\right],\ldots,A_{\nu_n}\right],
\end{eqnarray}
as can be seen inductively and therefore if $A_\nu$ is in the algebra,
so is $\Omega_{\nu_1\ldots \nu_n} \cdot A$.

\end{document}